\documentclass[electronic, preprint]{vgtc}                          

\graphicspath{{figures/}{pictures/}{images/}{./}} 

\usepackage{times}                     

\usepackage{tabu}                      
\usepackage{booktabs}                  
\usepackage{lipsum}                    
\usepackage{mwe}                       

\usepackage{mathptmx}                  
\usepackage{soul}
\usepackage{amsmath}
\usepackage{float}
\usepackage{fontawesome5}
\usepackage{graphicx}
\usepackage{subcaption}

\onlineid{0}

\vgtccategory{Research / Position}

\vgtcinsertpkg

\title{Are Average Estimates in Line Graphs Biased Toward Areas of Higher Variability?}

\author{Sheng Long\thanks{e-mail: shenglong2025@u.northwestern.edu} %
}
\affiliation{\scriptsize Northwestern University}

\abstract{
    When estimating the average value of a data series in line graphs, do viewers systematically overweight regions of higher variability? This is the question that Moritz et al.~\cite{moritz2023average} set out to answer, and across two preregistered experiments they reported evidence for a bias they termed variability overweighting. Using their released data and code, I first reproduce their reported analysis, then ask whether their inference survives an equally valid analytic choice: measuring estimation error against the average of the displayed marks of each stimulus, rather than against a reference series that, for one mark type, is never shown. Under this alternative operationalization, the effect is substantially reduced and, for one mark type, reverses. Splitting the analysis by whether variability appears in the upper or lower region of the graph further reveals an asymmetry that the original direction-corrected analysis collapses away. I argue that the deeper issue is what verbal theories leave unspecified. When a theory is silent about which operationalization it requires, the metric ends up doing the theory's arguing for it. I close with a discussion of what reproduction and reanalysis are for: not verdicts on truth, but instruments for making analytic contingency visible and for forcing theories to say more.
} 

\keywords{Graphical Perception, Reproduction, Inferential Reproducibility, Average Estimation}

\begin{document}


\firstsection{Introduction}

\maketitle

Estimating the average value of a data series is one of the most common tasks people perform when reading line graphs~\cite{albers2014task, correll2012comparing}. Understanding how properties of a visualization affect this estimation is central to graphical perception research and to the design guidelines that flow from it.

Moritz et al.~\cite{moritz2023average} investigated this question and reported a previously undocumented bias: average estimates in line graphs are systematically pulled toward areas of higher variability in the data, an effect they termed \textit{variability overweighting}. Across two preregistered experiments with 562 participants, they found that when one region of a line graph displayed more variable data, participants' average estimates shifted toward that region. In a second experiment comparing three mark types --- a continuous line, points equally spaced along the x-axis, and points equally spaced along the arc of the line --- the bias was smallest for x-axis-spaced points and largest for arc-spaced points. The authors interpret this pattern through \textit{visual salience}: marks that capture more attention should exert more pull on the estimate.


In prior work introducing visual decoding operators~\cite{long2026visual}, we used Moritz et al.'s second experiment as a task for our learned operators to generalize to. Using their experiment this way required reconstructing details of the stimulus generation and analysis pipeline from their published paper, supplementary code, and released data. During the course of reproducing the original analysis, we noticed that the original authors defined and computed estimation error differently from the operationalization we had assumed.\footnote{It is useful to be precise about terms, since ``reproduce'' and ``replicate'' carry several distinct meanings~\cite{jansen2024many,goodmanWhatDoesResearch2016}. This paper discusses the terms under the context of computational experiments. Following Jansen et al.~\cite{jansen2024many}, I use ``reproduce'' to mean that an independent group obtains the same result using the original authors' own artifacts, and ``replicate'' to mean that a different team, with a different experimental setup, obtains the same result. Goodman et al.~\cite{goodmanWhatDoesResearch2016} draw a finer distinction: \textit{methods reproducibility} (i.e., the same data and procedures yield the same numbers) and \textit{inferential reproducibility} (i.e., whether qualitatively similar conclusions are drawn from an independent replication or from a reanalysis of the original data). I find that Moritz et al.~\cite{moritz2023average} afford the former. \Cref{sec:recalc-error} probes the latter.}
When we recompute estimation error using an equally valid operationalization, the reported effect of variability overweighting is substantially reduced. Additionally, splitting the analysis by whether variability appears in the upper or lower region of the graph reveals an asymmetry that the original analysis did not address. Together, these observations suggest that the evidence for variability overweighting as a salience-driven effect is weaker than originally reported.

In the remainder of the paper, I review prior work on average estimation from visualizations, walk through the experimental design of Moritz et al., present the reproduction and reanalysis, and use the case to discuss the relationship between error definitions, verbal theories, and the conclusions we draw from graphical perception experiments.

\section{Estimating Averages from Visualizations}

Mean comparison and estimation has been studied extensively in data visualization. For example, Gleicher et al.~\cite{gleicherPerceptionAverageValue2013} examined relative mean judgments in multiclass scatterplots and found that viewers' performance is largely unaffected by the number of points per class, by redundant or conflicting secondary cues, or by additional distractor classes, but degrades when class membership is encoded through a less salient channel such as shape rather than color. Tseng et al.~\cite{tsengMeasuringCategoricalPerception2023} extended the same multiclass mean-estimation task to between two and ten categories under ten color palettes, and found that both the number of categories and the discriminability of colors within a palette significantly affect viewers' judgments. Beyond accuracy, Hong et al.~\cite{hongWeightedAverageIllusion2021} report a ``weighted average illusion'' in scatterplots that encode a third dimension through size or lightness of the dots. They found that larger and darker points are given more weight in x- and y-mean estimates, and the magnitude of the bias depends on the size or lightness range a designer maps onto the data. They account for the effect by modeling each mark's contribution and measuring attention to individual points using the centroid method from vision science. Correll et al.~\cite{correll2012comparing} asked participants to find the month with the highest average value in a year of daily data, and found that a colorfield encoding significantly outperformed the standard line graph on this aggregate judgment. For a broader treatment, a prior survey on ensemble perception~\cite{szafirFourTypesEnsemble2016} identified summary tasks, such as mean, skew, and numerosity, that are common to  data visualization and discusses related work. 

\section{Overview of Moritz et al.~\cite{moritz2023average}}

\begin{figure}[!htbp]
    \centering
    \includegraphics[width=\linewidth]{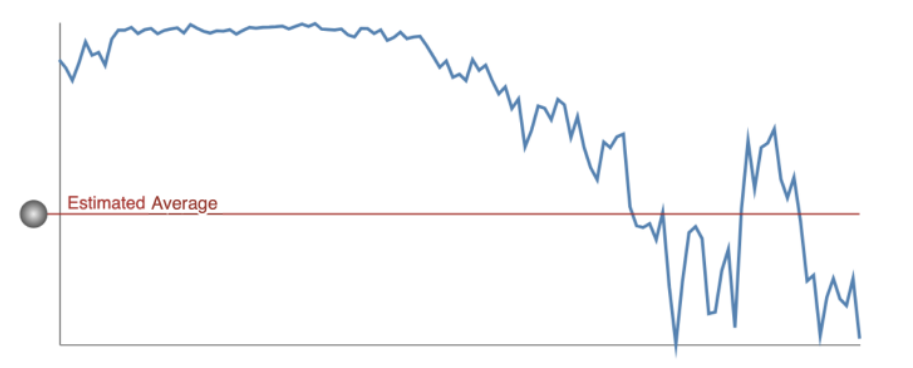}
    \caption{Screenshot of example stimulus shown in Moritz et al. Figure 3~\cite{moritz2023average}. Each series is shown as a 500px by 200px graph.}
    \label{fig:example-stimulus}
\end{figure}

Moritz et al.~\cite{moritz2023average} investigated what they term \textit{variability overweighting} in line graphs: the hypothesis that when estimating the average of a time series, viewers' judgments are systematically pulled toward regions of higher variability. They tested this hypothesis in two pre-registered experiments. In both experiments, stimuli were synthetic time series of data points generated from geometric Brownian motion, with noise linearly interpolated as a function of the base value, so that variability concentrates in the upper region of the graph or, after vertical mirroring, the lower region. Participants placed a draggable horizontal line where they judged the series' average to be (\Cref{fig:example-stimulus}). Moritz et al.~\cite{moritz2023average} release their experiment data and analysis files at \url{https://osf.io/aupbk/}.

The first experiment ($N = 142$) used line graphs only, in a $2$ (variability position: upper vs. lower) $\times 2$ (variability level: 0.15 vs. 0.4) within-subjects design. 12 seeds $\times$ 4 conditions gave 48 trials, each series containing 120 data points. Moritz et al. found that average estimates were biased toward the region of the line showing more variable data. 

From this finding, Moritz et al. hypothesized that the bias is driven by \textit{salience}\footnote{That is, ``how capturing [marks] are to our attention''} and that one ``could influence the degree of variability-overweighting by changing the mark type.'' 
Their second experiment ($N=420$) tests this with a mixed design: 3 mark types between subjects --- line graph, points equally spaced along the x-axis (\textsf{point}) and points equally spaced along the arc of the line (\textsf{pointArc}) --- and within subjects, 2 variability levels (0 vs 0.4) $\times$ 2 variability positions (upper vs lower) $\times$ $12$ random seeds, for $48$ trials per participant. The results in Experiment 2 replicated Experiment 1. Error increased with variability for every mark type, and the increase was smallest for \textsf{point}, larger for \textsf{line}, and largest for \textsf{pointArc}. Moritz et al. take this ordering as support for the salience account.


\section{Two Definitions of Estimation Error}
\label{sec:recalc-error}

\begin{figure}[!htbp]
    \centering
    \includegraphics[width=0.75\linewidth]{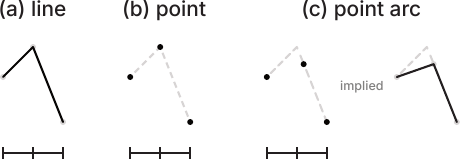}
    \caption{Given a line graph (\textsf{a}), a participant could be ``sampling'' from it using points equally spaced along the $x$-axis (\textsf{b}), or points equally spaced along the arc of the line (\textsf{c}).}
    \label{fig:diagram}
\end{figure}

Moritz et al.~\cite{moritz2023average} observe that in the \textsf{pointArc} condition, ``a perfectly accurate viewer cannot estimate the true average of the underlying data series''. This is correct only under one reading, when ``true average'' means the average of the time-series data that the dots were sampled from, rather than of the dots themselves (\Cref{fig:diagram}). Calling the unshown series the true average is therefore not a description of the encoding but a decision about what an estimate is measured against. Two definitions of error follow from the two readings, and I recalculate the same analysis under both definitions. 

I define \textit{displayed error} as the difference between a participant's reported average and the numeric average of the y-values of the marks actually presented in that stimulus. I define \textit{reference error} as the difference between a participant's reported average and the \textit{reference average}, that is, the average of the underlying data series from which all three mark types for a given seed, variability level, and variability position were generated.\footnote{This is what Moritz et al. call the ``true average'', and my definition of reference error is what they call ``error''. }

\begin{figure}[!htbp]
    \centering
    \includegraphics[width=0.95\linewidth]{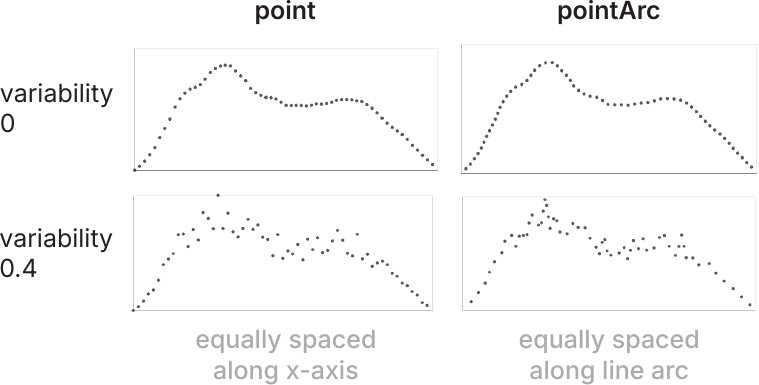}
    \caption{Example stimuli for the \textsf{point} and \textsf{pointArc} conditions, holding the variability position fixed to upper. 
    } 
    \label{fig:four-chart-example}
\end{figure}

For \textsf{line} and \textsf{point}, the displayed marks are the underlying series, so displayed error is reference error. For \textsf{pointArc} they do not: placing dots equally spaced along the line arc places more points in the high-variability region, so the displayed average sits closer to that region than the reference average does (see, e.g., \Cref{fig:four-chart-example}). 

For either quantity, the \textit{direction-corrected} version flips the sign so that positive values always indicate displacement toward the high-variability region: it equals the raw value when variability is in the upper region and its negation when variability is in the lower region. Therefore, \textit{direction-corrected reference error} is what Moritz et al. report as normalized error and plot in their Figure 6. Below I present results under both definitions. 


\subsection{Recomputing Estimation Error}

\begin{figure}[!hbtp]
    \centering
    \includegraphics[width=0.9\linewidth]{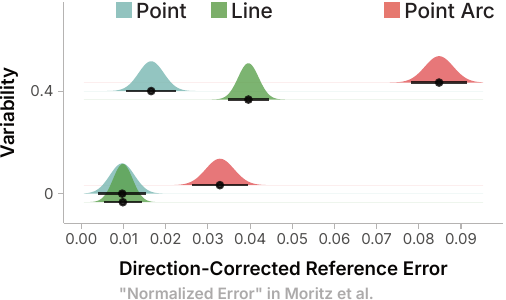}
    \caption{Reproducing Experiment 2 results presented in Moritz et al.~\cite{moritz2023average} Figure 6. The interval within each density plot shows the 95\% CI, with the black dot marking the mean.}
    \label{fig:og-recreate}
\end{figure}

In figure axes and captions, error without qualification means displayed error; reference error is always labeled as such. I reproduce Moritz et al.~\cite{moritz2023average}'s Figure 6 using the \texttt{SummarySEWithin} function from their supplementary materials (~\href{https://osf.io/aupbk/files/836ns}{\faFolderOpen~\texttt{Main\_Analysis/Functions.R}}~), plotting direction-corrected reference error by mark type and variability level (\Cref{fig:og-recreate}). The result matches the original.



\begin{figure}[!htbp]
    \centering
    \includegraphics[width=0.9\linewidth]{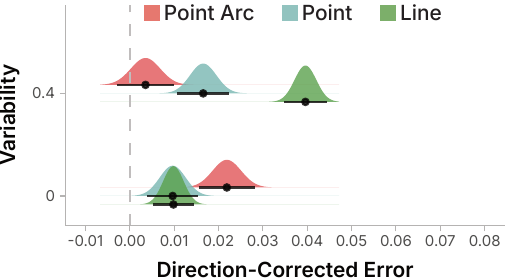}
    \caption{Recalculating Moritz et al. Experiment 2 results using the same analysis and plotting code as~\Cref{fig:og-recreate}, except with error instead of reference error.}
    \label{fig:newError}
\end{figure}


I follow the same procedure with direction-corrected displayed error in place of reference error (\Cref{fig:newError}). For \textsf{line} and \textsf{point}, nothing changes, as per definition. For \textsf{pointArc}, error at high variability drops to near zero, and the effect of added variability reverses sign. Participants, in other words, averaged the arc-spaced points about as accurately as they averaged anything. The distance between the point arc distributions in~\Cref{fig:og-recreate} is thus evidence of the distance between the displayed points and a series that was never shown, not a perceptual pull toward variability for line graphs consisted of dots.

\subsection{Error is Not Symmetric}

\begin{figure}[!htbp]
    \centering
    \includegraphics[width=0.9\linewidth]{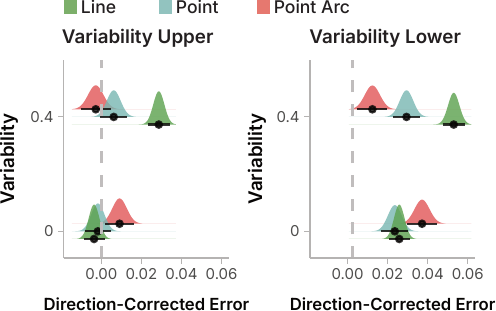}
    \caption{Direction-corrected error, split by whether variability appears in upper or lower y-values. Across mark types, conditions with variability in lower y-values show greater deviation from zero than their upper counterparts.}
    \label{fig:newError-Facet}
\end{figure}

Direction correction folds the upper- and lower-variability conditions into one. \Cref{fig:newError-Facet} unfolds them, plotting direction-corrected displayed error for each combination of mark type, variability level, and variability position.

I expected the two conditions to be similar; they are not. Across all three mark types, conditions with more variability in the lower y-values (``Variability Lower'') deviate further from zero than their upper counterparts, and the gap persists at variability 0, where the upper and lower stimuli differ only by vertical reflection. The released data cannot say whether the asymmetry comes from the probe's starting position, limits on how far it could be dragged, or the stimulus rescaling.  Future work could separate these accounts by varying the probe's starting position, adding un-mirrored controls, and logging the probe's reachable range.

\section{Discussion}
 
To summarize: what survives the change of metric is the comparison between the \textsf{point} and \textsf{line} conditions. Both display the same underlying data, so the two error definitions coincide there, and the original ordering --- \textsf{line} showing greater bias than \textsf{point} --- holds under either. What does not survive is the placement of \textsf{pointArc}, which depends entirely on which quantity is treated as ground truth, unless one is prepared to defend the auxiliary assumption that viewers of every mark type are estimating the same latent data series, in which case the reference average is the right ground truth after all. The salience account, as stated, neither asserts nor rules out that assumption.
 
\subsection{When the Theory is Silent, the Metric Speaks}
 
Moritz et al. motivate their error metric with a theoretical commitment: if variability overweighting is driven by visual salience, then the relevant ground truth is the average of the underlying data series, because salience is hypothesized to distort perception of that quantity regardless of mark type. This is a reasonable choice, but not a forced one: the salience account does not distinguish between the underlying data average and the displayed data average, so it cannot say which one error should be measured against. For the \textsf{line} and \textsf{point} conditions the distinction does not matter; for \textsf{pointArc} it does. The theory is silent precisely where the operationalization has consequences.
 
This is what Akbaba et al.~\cite{akbaba2024entanglements} call \textit{entanglement}: the theoretical commitment motivates the operationalization, and the operationalization produces the pattern that confirms the commitment --- an entanglement that stays invisible as long as only one operationalization is examined. In Fried's~\cite{fried2020lack,fried2020theories} terms, salience here operates as a \textit{weak theory}: a narrative account that specifies neither the functional form of the predicted relationship nor what would count as disconfirming evidence --- under a different metric, any result can be read as a mark type simply being ``less salient.'' A strong theory would specify enough that the choice of error metric is either derivable from the theory or explicitly flagged as a researcher degree of freedom.
 
This diagnosis suggests its own remedy. Guest and Martin~\cite{guest2021computational} argue that computational modeling forces theory building precisely because a formal model cannot remain silent: to implement a salience account of average estimation, one must commit to what salience is a function of, what quantity it distorts, and therefore which error metric measures the distortion. That is how the present observations arose: composing decoding operators to predict responses on Moritz et al.'s task~\cite{long2026visual} required committing to which quantity viewers are estimating --- a commitment the verbal account never has to make, and whose necessity became visible only when a generative model demanded it. A named effect without a specified mechanism is difficult to falsify, and what cannot be falsified cannot be killed~\cite{correll2018ross}. The field's growing catalog of named biases is not, in itself, progress; each name is a promissory note for a mechanism, and the note comes due when, as here, two reasonable operationalizations of the same claim disagree.
 
\subsection{What is Reproduction For? Where Should Explanation Stop?}
 
It is tempting to read a reanalysis like this one as a verdict: the effect ``failed to reproduce,'' therefore it is not real. That reading is too strong in both directions --- reproducibility and truth are separable properties: reproducible results need not be true, and irreproducible results need not be false~\cite{devezer2019scientific}. Nor am I claiming that visual salience is incorrect.
 
Every theory of the mind (and eye) buys tractability with assumptions; there is no free lunch. The question is where the assumptions live: whether they are explicit or implicit, and whether they were forced by modeling necessity or merely went unexamined. What troubles me about ``salience'' as an explanatory device is that it is where explanation stops. Once we arrive at ``more salient marks capture more attention,'' there is nothing beyond it --- not what salience is computed from, not how attention weights an estimate, not which quantity gets weighted. Every explanation stops somewhere; it matters where. Vision science stops later --- it maintains computational models of salience and formal accounts of ensemble perception --- and whether importing that machinery would sustain variability overweighting, dissolve it, or split it into several distinct effects is exactly the kind of question this workshop exists to ask.

\section*{Supplemental Materials}
\label{sec:supplemental_materials}

All supplemental materials are available on OSF at \url{https://osf.io/g73wc/}, released under a CC BY 4.0 license.

\section*{Figure Credits}
\label{sec:figure_credits}

\Cref{fig:example-stimulus} is a screenshot of Fig.\ 3 in Moritz et al and \Cref{fig:og-recreate} is a recreation of Fig.\ 6 from Moritz. et al.~\cite{moritz2023average}, which is in the public domain. The remaining figures (\cref{fig:newError,fig:newError-Facet}) followed \Cref{fig:og-recreate} in style. The first author created the diagram in \Cref{fig:diagram}. 



\acknowledgments{
I thank the reviewers for their helpful feedback and Matthew Kay for early discussion of this work and the authors of Moritz et al.~\cite{moritz2023average} for publicly releasing their data and analysis code in an easily accessible form. 
}

\bibliographystyle{abbrv-doi-hyperref-narrow}

\bibliography{template}
\end{document}